\documentclass[12pt,twoside,aps,pra,nofootinbib,amsmath,amssymb,tightenlines,floats,floatfix]{revtex4}
\usepackage{epsfig}
\usepackage{psfig}
\usepackage{mathptmx}
\usepackage{bm}
\usepackage{graphicx}
\tighten
\begin{document}
\title{Scattering length of the helium atom -- helium dimer collision}
\author{Elena A. Kolganova, Alexander K. Motovilov}
\address{Bogoliubov Laboratory of Theoretical Physics,
Joint Institute  for Nuclear Research\\ Joliot-Curie 6, 141980 Dubna, Moscow Region, Russia}
\author{Werner Sandhas}
\address{Physikalisches Institut der Universit\"at Bonn\\
Endenicher Allee 11-13, D-53115 Bonn, Germany}

\date{August 04, 2004} 

\begin{abstract}
We present our recent results on the scattering length
of  $^4$He--$^4$He$_2$ collisions. These investigations
are based on the hard-core version of the Faddeev
differential equations. As compared to our previous
calculations of the same quantity, a much more refined
grid is employed, providing an improvement of
about 10\%. Our results are compared with other {\it ab
initio}, and with model calculations.
\\\\
{PACS numbers (2001): 21.45.+v, 34.50.-s, 02.60.Nm}
\end{abstract}

\maketitle

\section{Introduction}
\label{SIntro}

Weakly bound small $^4$He clusters attracted
considerable attention in recent years, in particular
because of the booming interest in Bose-Einstein
condensation of ultra-cold gases
\cite{BECTheory,molBEC}.

Experimentally, helium dimers have been observed in
1993 by Luo {\em at al.} \cite{DimerExp}, and in 1994
by Sch\"ollkopf and Toennies \cite{Science}.
In the latter investigation the existence of helium
trimers has also been demonstrated. Later on, Grisenti {\em
et al.} \cite{exp} measured a bond length of $52 \pm 4$
{\AA} for $^4$He$_2$, which indicates that this dimer
is the largest known diatomic molecular ground state.
Based on this measurement they estimated a scattering
length of $104^{+8}_{-18}$ {\AA} and a dimer energy of
$1.1^{+0.3}_{-0.2}$ mK \cite{exp}. Further
investigations concerning helium trimers and tetramers
have been reported in
Refs. \cite{Toennies1996,Toennies2002}, but with no
results on size and binding energies.

Many theoretical calculations of these systems were
performed for various interatomic potentials
\cite{Aziz87,Aziz91,TTY,Aziz97}. Variational,
hyperspherical and Faddeev-type techniques have been
employed in this context (see, e.g.,
\cite{Nakai}--\cite{RoudnevE} and references therein).
For the potentials given in \cite{Aziz91,TTY} it turned
out that the Helium trimer has two bound states of
total angular momentum zero: a ground state of about
$126$ mK and an excited state of about $2.28$ mK. The
latter was shown to be of Efimov
nature~\cite{Gloeckle,EsryLinGreene,KM-YAF}. In
particular, it was demonstrated in~\cite{KM-YAF} how
the Efimov states emerge from the virtual ones when
decreasing the strength of the interaction. High
accuracy has been achieved in all these calculations.

While the number of papers devoted to the $^4$He$_3$
bound-state problem is rather large, the number of
scattering results is still very limited. Phase shifts
of $^4$He--$^4$He$_2$ elastic scattering at ultra-low
energies have been calculated for the first time in
\cite{KMS-PRA,MSK-CPL} below and above the three-body
threshold. An extension and numerical improvement of
these calculations was published in \cite{MSSK}. To the
best of our knowledge, the only alternative {\em ab
initio} calculation of phase shifts below the
three-body threshold was performed in \cite{RoudnevE}.
As shown in \cite{BHvK,BraatenHammer}, a zero-range
model formulated in field theoretical terms is able to
simulate the scattering situation.

Though being an ideal quantum mechanical problem,
involving three neutral bosons without complications
due to spin, isospin or Coulomb forces, the exact
treatment of the $^4$He triatomic system is numerically
quite demanding at the scattering threshold. Due to the
low energy of the Helium dimer, a very large domain  in
configuration space, with a characteristic size of
hundreds of {\AA}ngstroems, has to be considered. As a
consequence, the accuracy achieved in
\cite{KMS-JPB,MSSK} for the scattering length appeared
somewhat limited. To overcome this limitation, we have
enlarged in the present investigation the cut-off
radius $\rho_{\rm max}$ from 600 to 900 {\AA} and
employed much more refined grids.

\section{Formalism}

Besides the complications related to the large domain
in configuration space, the other source
of complications is the strong repulsion of the He--He
interaction at short distances. This problem, however,
was and is overcome in our previous and present
investigations by employing the rigorous hard-core
version of the Faddeev differential equations developed
in \cite{Vestnik,MerMot}.

Let us recall the main aspects of the corresponding
formalism (for details see \cite{KMS-JPB,MSSK}).
In what follows we restrict ourselves to a total
angular momentum $L=0$. In this case one has to solve
the two-dimensional integro-differential
Faddeev equations
\begin{equation}
\label{FadPart}
   \left[-\displaystyle\frac{\partial^2}{\partial x^2}
            -\displaystyle\frac{\partial^2}{\partial y^2}
            +l(l+1)\left(\displaystyle\frac{1}{x^2}
            +\displaystyle\frac{1}{y^2}\right)
    -E\right]\Phi_l(x,y)=\left\{
            \begin{array}{cl} -V(x)\Psi_l(x,y), & x>c \\
                    0,                  & x<c\,.
\end{array}\right.
\end{equation}
Here, $x,y$ stand for the standard Jacobi variables and
$c$ for the core range.  The angular momentum  $l$
corresponds to a dimer subsystem and a complementary
atom; for an $S$-wave three-boson state, $l$ is even
($l=0,2,4,\ldots\,).$ $V(x)$ is the He-He central
potential acting outside the core domain. The partial
wave function $\Psi_l(x,y)$ is related to the Faddeev
components $\Phi_l(x,y)$ by
\begin{equation}
\label{FTconn}
         \Psi_l(x,y)=\Phi_l(x,y) + \sum_{l'}\int_{-1}^{+1}
         d\eta\,h_{l l'}(x,y,\eta)\,\Phi_{l'}(x',y'),
\end{equation}
where
$$
          x'=\sqrt{\displaystyle\frac{1}{4}\,x^2+\displaystyle
    \frac{3}{4}\,y^2-\displaystyle\frac{\sqrt{3}}{2}\,xy\eta}\,,
\qquad
         y'=\sqrt{\displaystyle\frac{3}{4}\,x^2+\displaystyle
   \frac{1}{4}\,y^2+ \displaystyle\frac{\sqrt{3}}{2}\,xy\eta}\,,
$$
and  $1 \leq{\eta}\leq 1$. The explicit form of the function
$h_{ll'}$ can be found in Refs.~\cite{MF,MGL}.

The functions $\Phi_{l}(x,y)$ satisfy the boundary conditions
\begin{equation}
\label{BCStandard}
      \Phi_{l}(x,y)\left.\right|_{x=0}
      =\Phi_{l}(x,y)\left.\right|_{y=0}=0\,.
\end{equation}
Moreover, in the hard-core model they are required to
satisfy the condition
\begin{equation}
\label{BCCorePart}
       \Phi_{l}(c,y) + \sum_{l'}\int_{-1}^{+1}
       d\eta\,h_{l l'}(c,y,\eta)\,\Phi_{l'}(x',y')=0\,.
\end{equation}
This guarantees the wave function  $\Psi_{l}(x,y)$ to be zero not only at the
core boundary $x=c$ but also inside the core domains.

The asymptotic boundary condition for the partial wave Faddeev
components of the two-fragment scattering
states reads, as $\rho\rightarrow\infty$ and/or
$y\rightarrow\infty$,
\begin{equation}
\label{AsBCPartS}
    \begin{array}{rcl}
      \Phi_l(x,y;p) & = &
      \delta_{l0}\psi_d(x)\left\{\sin(py) + \exp({\rm i}py)
      \left[{\rm a}_0(p)+o\left(y^{-1/2}\right)\right]\right\} \\
      && +
  \displaystyle\frac{\exp({\rm i}\sqrt{E}\rho)}{\sqrt{\rho}}
                \left[A_l(\theta)+o\left(\rho^{-1/2}\right)\right].
    \end{array}
\end{equation}
Here, $\psi_d(x)$ is  the dimer wave function, $E$
stands  for the scattering energy given by
$E=\varepsilon_d+p^2$ with $\varepsilon_d$ the dimer energy,
and $p$ for the relative momentum conjugate to the
variable $y$. The variables $\rho=\sqrt{x^2+y^2}$ and
$\theta=\arctan\displaystyle\frac{y}{x}$ are the
hyperradius and hyperangle, respectively. The
coefficient ${\rm a}_0(p)$ is nothing but the elastic
scattering amplitude, while the functions $A_l(\theta)$
provide us, at $E>0$, with the corresponding
partial-wave Faddeev breakup amplitudes. The $^4$He --
$^4$He$_2$ scattering length $\ell_{\rm sc}$ is given
by
\begin{equation}
\label{sclen}
\ell_{\rm sc}=-\displaystyle\frac{\sqrt{3}}{2}\,
\begin{array}{c}\phantom{a}\\
{\rm lim}\,\\
\mbox{\scriptsize$p\rightarrow0$}
\end{array}\,
\frac{{\rm a}_0(p)}{p}.
\end{equation}

Surely we only deal with a finite number of equations
(\ref{FadPart})--(\ref{BCCorePart}), assuming
$l\leq l_{\rm max}$, where $l_{\rm max}$ is a certain
fixed even number. As in \cite{KMS-JPB,MSSK} we use a
finite-difference approximation of the boundary-value
problem (\ref{FadPart})--(\ref{AsBCPartS}) in the polar
coordinates $\rho$ and $\theta$. The grids are chosen
such that the points of intersection of the arcs
$\rho=\rho_i$, $i=1,2,\ldots, N_\rho$ and the rays
$\theta=\theta_j$, $j=1,2,\ldots, N_\theta$ with the
core boundary $x=c$ constitute the  knots. The value of
the core radius is chosen to be $c=1$\,{\AA} by the
argument given in \cite{MSSK}. We also follow the same
method for choosing the grid radii $\rho_i$ (and, thus,
the grid hyperangles $\theta_j$) as described in
\cite{KMS-JPB,MSSK}.


\section{Results}

Our calculations are based on the semi-empirical HFD-B
\cite{Aziz87} and LM2M2 \cite{Aziz91} potentials by
Aziz and co-workers, and the more recent, purely
theoretically derived TTY \cite{TTY} potential by Tang,
Toennies and Yiu. For the explicit form of these
polarization potentials we refer to the Appendix of
Ref. \cite{MSSK}. As in our previous calculations we
choose $\hbar^2/m=12.12$\,K\,\AA$^2$, where $m$ stands
for the mass of the $^4$He atom. The $^4$He dimer
binding energies and $^4$He--$^4$He
scattering lengths obtained with the HFD-B, LM2M2, and
TTY potentials are shown in Table \ref{T2body}. Note
that the inverse of the wave number
$\varkappa^{(2)}=\sqrt{|\varepsilon_d|}$ lies rather close
to the corresponding scattering length.


\begin{table}[hb]
\caption{Dimer energy $\varepsilon_d$, wave
length $1/\varkappa^{(2)}$, and
$^4$He$-$$^4$He scattering length $\ell_{\rm sc}^{(2)}$
for the potentials used, as compared to the experimatal
values of Ref. \cite{exp}.} \label{tableDimerLen}
\begin{center}
\begin{tabular}{|ccc|cccc|}
\hline
 & \quad $\varepsilon_d$ (mK)\quad & \quad $\ell^{(2)}_{\rm sc}$ (\AA)\quad &
{\quad Potential\quad} & \quad $\varepsilon_d$ (mK)\quad & \quad $1/\varkappa^{(2)}$ (\AA)\quad
& \quad $\ell^{(2)}_{\rm sc}$ (\AA) \quad \\
\hline
  &   &  &  LM2M2 & $-1.30348$ & 96.43 & 100.23 \\
\quad Exp. \cite{exp}\quad & $1.1^{+0.3}_{-0.2}$  &$104^{+8}_{-18}$ & TTY   & $-1.30962$ & 96.20 & 100.01 \\
 &   & & HFD-B  & $-1.68541$ &  84.80 & $ 88.50$ \\
\hline
\end{tabular}
\label{T2body}
\end{center}
\end{table}

\begin{table}[h]
\caption {The $^4$He--$^4$He$_2$ scattering length
$\ell_{\rm sc}$ (\AA) for $\ell_{\rm max}=0$ in case of
the TTY potential as a function of the grid parameters
$\rho_{\rm max}$ and $N=N_\rho=N_\theta$.}
\label{TTYLen_conv}
\begin{center}
\begin{tabular}{lcccccc}
\hline
& & & & &  \\[-2ex]
\phantom{aaaaaaa} $N$ & {\quad 1005 \quad} & {\quad 1505 \quad} &
{\quad 2005 \quad} & {\quad 2505 \quad} & {\quad 3005 \quad} & {\quad 3505 \quad} \\[-2ex]
 $\rho_{\rm max}$ &      & & &  & &\\
 & & & & &  &\\[-2ex]
\hline
& & & & & & \\[-2ex]
  600 &     162.33 &159.80 & 158.91 &158.61   &158.31& \\
  700 & 164.13 &159.99 &158.57&157.99 & 157.65 & 157.48\\
  800 & 167.15 &160.98 & 158.90 & 158.03&157.46 & \\
 900 &171.19  & 162.52&159.66&158.40&  157.66 & \\
\hline
\end{tabular}
\end{center}
\end{table}

\begin{table}[htb]
\caption{The $^{4}$He--$^4$He$_2$ scattering length
$\ell_{\rm sc}$ ({\AA}) obtained for a grid with
$N_\rho=N_\theta$=2005 and $\rho_{\rm max}$=700~\AA. }
\label{tableScLength}
\begin{center}
\begin{tabular}{ccccccccc}
\hline
{\quad Potential\quad}  & {\quad $l_{\rm max}$\quad} &
{\quad This work \quad}& {\quad \cite{MSSK}\quad}  & {\quad \cite{BlumeGreene}\quad}
& {\quad \cite{RoudnevE}\quad} & {\quad\cite{Penkov} \quad} &
{\quad \cite{BraatenHammer} \quad} & \\
\cline{1-2}\cline{3-9}
       &  0 & 158.2& $168$  &     &       &&& \\
 LM2M2 &  2 & 122.9 & $134$  &     &       &&& \\
       &  4 & 118.7 & $131$  &  126  & 115.4 &114.25&113.1& \\
\cline{1-2}\cline{3-9}
       &  0 & 158.6& $168$   &     &        &&&  \\
 TTY   &  2 & 123.2& $134$   &     &        &&& \\
       &  4 & 118.9& $131$   &     & 115.8  &&114.5& \\
\cline{1-2}\cline{3-9}
       &  0 & 159.6& $168$   &     &        &&& \\
 HFD-B &  2 & 128.4& $138$   &     &        &&& \\
       &  4 & 124.7& $135$   &     & 121.9  &&120.2& \\
\hline
\end{tabular}
\end{center}
\end{table}

\begin{figure}[h]

\vspace*{1truecm}
\hspace*{-12.3truecm}$\ell_{\rm sc}$ ({\AA})
\vspace*{-0.5truecm}

\hspace*{10truecm}\begin{tabular}{c}
$\rho_{\rm max}=600$ {\AA}\\[-0.6mm]
$\rho_{\rm max}=700$ {\AA}\\[-0.6mm]
$\rho_{\rm max}=800$ {\AA}\\[-0.6mm]
$\rho_{\rm max}=900$ {\AA}
\end{tabular}

\vspace*{-4.4truecm}

\begin{center}
{\hspace{-1.5cm}\includegraphics[angle=-90,width=14.cm]{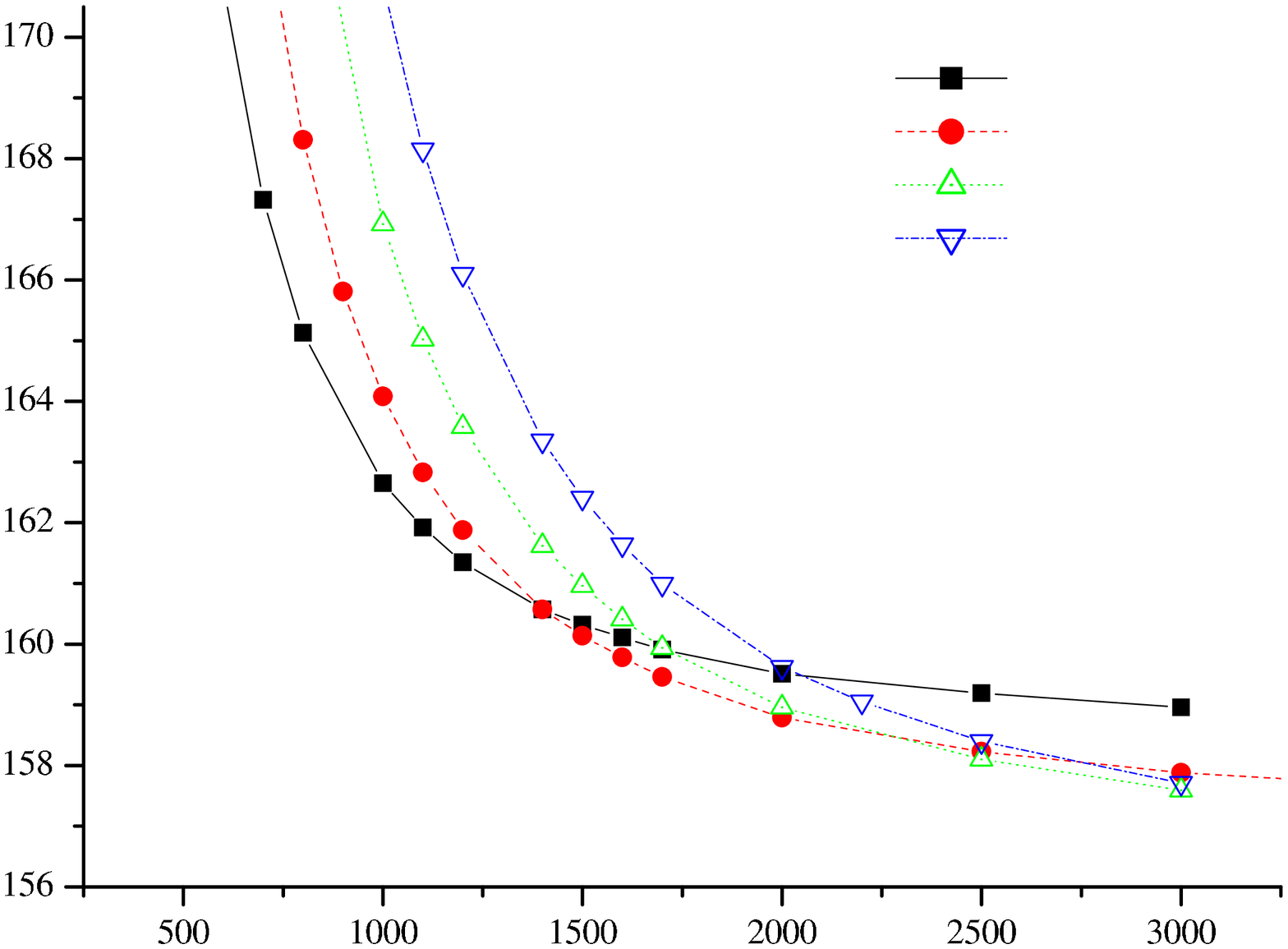}}
\end{center}

\vspace*{-0.2truecm}

\hspace*{11.truecm} $N$

\caption{\small The $^4$He--$^4$He$_2$ scattering length
$\ell_{\rm sc}$ for $l_{\rm max}=0$ in case of the TTY
potential as a function of the grid
dimension $N \equiv N_\rho=N_\theta$.}
\label{Fig-TTYLen}
\end{figure}

Unlike the trimer binding energies, the
$^4$He--$^4$He$_2$ scattering length is much more
sensitive to the grid parameters. To investigate this
sensitivity, we take increasing values of the cut-off
hyperradius $\rho_{\rm max}$, and simultaneously
increase the dimension of the grid $N=N_\theta=N_\rho$.
Surely, in such an analysis we can restrict ourselves
to $l_{\rm max}=0$. The results obtained for the TTY
potential are given in Table \ref{TTYLen_conv} and Fig.
\ref{Fig-TTYLen}. Inspection of this figure shows that,
when increasing the dimension $N$ of the grid,
convergence of the $^4$He--$^4$He$_2$ scattering length
$\ell_{\rm sc}$ is essentially achieved, however, with
different limiting values of  $\ell_{\rm sc}$ for
different choices of $\rho_{\rm max}$. This concerns,
in particular, the transition from $\rho_{\rm max}=600$
{\AA} to $\rho_{\rm max}=700$ {\AA}, while the
transition to $800$ {\AA} or even $900$ {\AA} has
practically no effect.

Bearing this in mind, we feel justified to
choose $\rho_{\rm max}=700$ {\AA} when going over from
$l_{\rm max} = 0$ to $l_{\rm max} = 2$ and 4.
The corresponding results are presented in Table
\ref{tableScLength}.
There we also show our previous results \cite{MSSK}
where, due to lack of computer facilities, we had to
restrict ourselves to $\rho_{\rm max}=460$ {\AA} and
$N=605$. We see that an improvement of about 10\% is
achieved in the present calculations, as indicated
already by the trends in Fig. \ref{Fig-TTYLen}.

Table \ref{tableScLength} also contains the fairly
recent results by Blume and Greene \cite{BlumeGreene}
and Roudnev \cite{RoudnevE}. The treatment of
\cite{BlumeGreene} is based on a combination of the
Monte Carlo method and the hyperspherical adiabatic
approach. The one of Ref. \cite{RoudnevE} employs the
three-dimensional Faddeev differential equations in the
total angular momentum representation. Our results
agree rather well with these alternative calculations.

This gives already a good hint on the quality
of our present investigations. A direct
confirmation is obtained by extrapolating the curves in
Fig. \ref{Fig-TTYLen}. According to this figure,
convergence of $\ell_{\rm sc}$ as a function of $N$ is
essentially, but not fully, achieved. A certain
improvement, thus, is still to be expected when going
to higher $N$. In order to estimate this effect we
approximate the curves of Fig. \ref{Fig-TTYLen} by a
function of the form
\begin{equation}
\label{extrap}
\ell_{\rm sc}(N)=\alpha+\frac{\beta}{N-\gamma}.
\end{equation}
Clearly, $\ell_{\rm sc}(\infty)=\alpha$. The constants
$\alpha$, $\beta$, and $\gamma$ are fixed by the values
of $\ell_{\rm sc}$ at $N=1005$, 2005, and $3005$. In
this way we get the corresponding optimal scattering
lengths $\ell_{\rm sc}(\infty)=157.5$, 156.4, 155.4,
and 154.8 {\AA} for $\rho_{\rm max}=600$, 700, 800, and
900 {\AA}, respectively. Comparing with Table
\ref{TTYLen_conv} shows that the differences between
these asymptotic values and the ones for $N=3005$ lie
between 1 to 3 {\AA}.

For $l_{\rm max} = 4$, $\rho_{\rm max}=700$ {\AA} and
the LM2M2 potential the scattering length has been
calculated for $N$ = 1005, 1505, and 2005. Employing
again the extrapolation formula (\ref{extrap}) with
$\alpha$, $\beta$, $\gamma$ being chosen according to
these values, we find $\ell_{\rm sc}(\infty)$ = 117.0
{\AA}. The difference between the scattering length
obtained for $N=2005$ and the extrapolated value,
hence, is 1.7 {\AA}. A direct calculation for higher
$N$ should lead to a modification rather close to this
result. Following this argumentation, we conclude that
the true value of $\ell_{\rm sc}$ for the LM2M2 and TTY
potentials lies between 115 and 116 {\AA}.

For completeness we mention that besides the above {\em
ab initio} calculations there are also model
calculations, the results of which are given in the
last two columns of Table \ref{tableScLength}. The
calculations of \cite{Penkov} are based on employing a
Yamaguchi potential that leads to an easily solvable
one-dimensional integral equation in momentum space.
The approach of \cite{BraatenHammer} (see also
\cite{BHvK} and references therein) represents
intrinsically a zero-range model with a cut-off
introduced to make the resulting one-dimensional
Skornyakov-Ter-Martirosian equation \cite{STM} well
defined. The cut-off parameter in
\cite{BHvK,BraatenHammer} as well as the range
parameter of the Yamaguchi potential in \cite{Penkov}
are adjusted to the three-body binding energy obtained
in the {\it ab initio} calculations. In other words, these
approaches are characterized by a remarkable
simplicity, but rely essentially on results of the {\it
ab initio} three-body calculations.


\acknowledgements We are indebted to
Prof.~V.\,B.\,Belyaev and Prof.~H.\,Toki for providing
us with the possibility to perform calculations at the
supercomputer of the Research Center for Nuclear
Physics of Osaka University, Japan. This work was
supported by the Deu\-t\-s\-che
For\-s\-ch\-ungs\-ge\-mein\-schaft (DFG), the
Heisenberg-Landau Program, and the Russian Foundation
for Basic Research.



\end{document}